\documentstyle[12pt]{article}
\begin{document}

\pagestyle{plain}

\vspace*{-10mm}

\baselineskip18pt
\begin{flushright}
{\bf BROWN-HET-968}\\
{\bf BROWN- TA-516}\\
{\bf SNUTP - 94 - 97}\\
{\bf October 1994}\\
\end{flushright}
\vspace{1.0cm}
\begin{center}

{\Large \bf Electroweak Physics at LEP\footnote{Presented
at the Workshop on Quantum Infrared Physics, American
University of Paris (Paris, France, 6 - 10 June, 1994)}}\\

\vglue 5mm
{\bf Kyungsik Kang } \\
\vglue 2mm
{\it Department of Physics, Brown University
, Providence, RI, USA\footnote{Permanent address and supported
in part by the USDOE contract DE-FG02-91ER40688-Task A} }\\
\vglue 1mm
{\it and}\\
\vglue 1mm
{\it LPTPE, Universite P.\& M. Curie, 4 Place Jussieu, 75252 Paris
Cedex05, France }\\
\vglue 3mm
{\it and}\\
\vglue 3mm
{\bf Sin Kyu Kang } \\
\vglue 2mm

{\it Department of Physics, Seoul National University, Seoul,
Korea\footnote{Supported in part by the Basic Science Research
Institute Program, Ministry of Education, Project No. BSRI-94-2418,
and the Korea Science and Engineering Foundation through SNU CTP} }\\
\vglue 10mm
{\bf ABSTRACT} \\
\vglue 8mm
\begin{minipage}{14cm}
{\normalsize
We have examined the evidence for the electroweak radiative
corrections in the LEP precision data sets of 1993 and 1994
along with the intriguing
possibility that the QED corrections only may be sufficient
to fit the data within the framework of the minimal standard model.
We find that the situation is very sensitive to the precise value
of $M_W$. The current world average value of $M_W$
and the improved 1994 LEP data strongly favor nonvanishing
electroweak radiative corrections,
and are consistent with a heavy $m_t$ as reported by CDF
but with a heavy Higgs scalar of about 400 GeV.
We discuss how future precision measurements of $M_W$ and $m_t$ can
provide a decisive
test for the standard model with radiative corrections.

}
\end{minipage}
\end{center}
\newpage

Recently much interests have been paid to the electroweak radiative
corrections (EWRC)
and precision tests of the standard model (SM) thanks to the accurate data
obtained at LEP [1,2]. Numerous articles have appeared on the subject
as has been documented
in [1 - 3]. The LEP data are generally   regarded as the success of the SM
and as the  evidence  for the nonvanishing EWRC [4].

There have been new experimental developments since last year that warrant
a renewed examination of the precision tests of the SM, namely, the new
measurements of $M_W$ [5], the improved LEP precision data [2],
and the evidence of $m_t$ from CDF [6].
We would like to report on the results of the new precision tests of the SM
based on these new experimental informations and implication on
the Higgs mass range. At the same time we reexamine
the intriguing claim made by
Novikov, Okun, and Vysotsky [7]
based on the 1993 experimental
data from LEP that the electroweak parameters as defined in the SM could be
explained by the QED Born approximation (QBA) in which $\alpha (M^2_Z)$
is used
instead of $\alpha(0)$ and the corresponding redefinition of the weak
mixing angle $\sin ^2\theta $ instead of $\sin ^2\theta_W $ in the tree-level
SM within 1 $\sigma$ level.

The full one-loop EWRC are calculated with the aid of the ZFITTER
program [8] modified by the improved QCD correction factor
and the $\chi^2$ minimization to the fit. In order to achieve QBA,
we neglect the terms of non-photonic and pure weak interaction origin
systematically in the program.

Since the basic lagrangian contains the bare electric charge $e_0$,
the renormalized physical charge $e$ is fixed by a counter term $\delta e$;
  $e_0 = e + \delta e $.
The counter term $\delta e$ is determined by the condition of the on-shell
charge renormalization in the $\overline{MS}$ or on-shell scheme.
It is well known that the charge renormalization in the conventional QED
fixes the counter term by
the renormalized vacuum polarization $\hat {\Pi}^{\gamma }(0)$ and
one can evaluate
$\hat {\Pi}^{\gamma}(q^2)=\hat {\Sigma} ^{\gamma \gamma }(q^2)/q^2$
from the photon self energy $\hat{\Sigma}^{\gamma \gamma }(q^2)$, for example,
by the dimensional regularization method. This gives at $q^2=M_Z^2$,
\begin{equation}
  \hat {\Pi}^{\gamma}(M_Z^2) = \sum_{f} Q_f^2\frac{\alpha}{3\pi}
  \left (\frac{5}{3}-\ln {\frac{M_Z^2}{m_f^2}}+i\pi \right),
\end{equation}
where $Q_f$ is the charge of the fermion $f$ in the unit of $e$ and
$\alpha$ is the hyperfine structure constant
$\alpha = \frac{e^2}{4\pi} = 1/137.0359895(61) $.

This gives the total fermionic contribution of $m_f \leq M_Z $ to the real
part,
 $Re\hat {\Pi}^{\gamma}(M_Z^2) = -0.0602(9) $,
so that the "running" charge defined as
$e^2(q^2) = \frac{e^2}{1+Re\hat {\Pi }^{\gamma}(q^2)} $
gives $\alpha(M_Z^2)=1/128.786 $ in the on-shell scheme.
The concept of the running charge is, however, scheme dependent [9]: the
$\overline {MS}$ fine structure constant at the $Z$ mass scale is given by
\begin{equation}
   \hat{\alpha }(M_Z)=\alpha /[1-\Pi^{\gamma}(0)|_{\overline {MS}}+
  2\tan {\theta_W}(\Sigma^{\gamma Z}(0)/M^2_Z)_{\overline{MS}}].
\end{equation}
so that one can show
                $\hat{\alpha}(M_Z) = (127.9\pm 0.1)^{-1}$,
which  differs by some 0.8 $\%$ from the on-shell $\alpha(M_Z^2)$.

The electroweak parameters are evaluated
numerically with the hyperfine structure constant $\alpha$,
the four-fermion coupling constant of $\mu$-decay,
$G_{\mu} &=& 1.16639(2)\times 10^{-5} ~~\mbox{GeV}^{-2}$,
and $Z$-mass, i.e., $M_Z &=& 91.187(7) ~~\mbox{GeV}$ for the 1993 data fit
and $91.1899(44) ~~\mbox{GeV}$ for the 1994 data fit.
Numerical estimate of the full EWRC requires
the mass values of the leptons, quarks, Higgs scalar  and $W$-boson besides
these quantities.
While $Z$-mass is known to an incredible accuracy from the LEP experiments
largely due to the resonant depolarization method, the situation with
respect to the $W$-mass is desired to be improved, i.e.,
$M_W = 80.22(26)~ $ GeV [10] and
$80.21(16)~ $ GeV [5] vs.
the     CDF measurement $M_W = 79.91(39)~ $ GeV [11] and
$80.38(23)~ $ GeV [5].

One has, in the standard model, the on-shell relation
$\sin^2 \theta_W = 1-\frac{M_W^2}{M_Z^2} $,
and the four-fermion coupling constant $G_{\mu}$
\begin{equation}
G_{\mu } = \frac{\pi \alpha }{\sqrt{2}M_W^2}
\left(1-\frac{M_W^2}{M_Z^2}\right)^{-1}(1-\Delta r)^{-1}
\end{equation}
so that $\Delta r$, representing the radiative corrections, is
given by
\begin{equation}
\Delta r = 1-\left(\frac{37.28}{M_W}\right)^2
\frac{1}{1-M_W^2/M_Z^2}.
\end{equation}

Notice that the radiative correction $\Delta r$ is
very sensitive
to the value of $M_W$:
Mere change in $M_W$ by $0.59\%$ results as much as a $75\%$ change in
$\Delta r$. Also precise determination of the on-shell value of $\sin
^2\theta_W$ can constrain the needed radiative correction and the value of
$M_W$, thus providing another crucial test for the evidence of the EWRC
in the standard model.

We have made $\chi^2$-fits to both 1993 and 1994 data sets
of the Z-decay parameters measured at LEP and $M_W$ as shown in Tables 1 - 4.
In each set, the fit is carried out for both the CDF and
world average values of $M_W$. Details of the analysis can be found
elsewhere [12].

The Z-decay parameters are
calculated with a modified ZFITTER program, in which
the best $\chi^2$ fit search is made
with the gluonic coupling constant
$\bar{\alpha _s}(M_Z^2) = 0.123 \pm 0.006$
in the improved QCD correction factor [13]
$R_{\mbox{QCD}} = 1+1.05\frac{\bar{\alpha_s}}{\pi}+0.9(\pm 0.1)
\left(\frac{\bar{\alpha_s}}{\pi}\right)^2-13.0
\left(\frac{\bar{\alpha_s}}{\pi}\right)^3$
for all quarks. The partial width for $Z\rightarrow f\bar{f}$ is given by
\begin{equation}
\Gamma_f = \frac{G_{\mu}}{\sqrt{2}}\frac{M_Z^3}{24\pi}\beta R_{\mbox{QED}}
c_fR_{\mbox{QCD}}(M_Z^2)\left \{ [(\bar{v}^Z_f)^2+(\bar{a}^Z_f)^2]\times
 \left(1+2\frac{m_f^2}{M_Z^2}\right)-6(\bar{a}^Z_f)^2\frac{m_f^2}
{M_Z^2}\right \}
\end{equation}
where $\beta =\beta(s)=\sqrt{1-4m_f^2/s}$ at $s=M_Z^2$, $R_{\mbox{QED}}
=1+\frac{3}{4}\frac{\alpha}{\pi}Q_f^2$ and the color factor $c_f=3$ for
quarks and 1 for leptons.
Here the renormalized vector and axial-vector couplings are defined by
$\bar{a}_f^Z=\sqrt{\rho_f^Z}2a_f^Z = \sqrt{\rho_f^Z}2I_3^f $ and
$\bar{v}^Z_f=\bar{a}^Z_f[1-4|Q_f|\sin^2\theta_W\kappa^Z_f] $ in
terms of the familiar notations [8,9,14].
Note that $\Delta \alpha $  is contained in the
couplings through $G_{\mu}$
and all other
non-photonic and pure weak loop corrections are
grouped in $\rho_f^Z$ and $\kappa_f^Z$.
Thus the case of the QBA
can be achieved simply by setting $\rho^Z_f$ and $\kappa^Z_f$
to 1 in the vector and axial-vector couplings.

\begin{table}
\begin{center}
\begin{tabular}{|c||c|c||c|c|c|} \hline \hline

 & Experiment & QBA & Full EW & Full EW & Full EW \\
  \hline
 $m_t$~(GeV) & 150  & & 120 & 138 & 158 \\
 $ m_H$~(GeV) & 60 $\leq m_H \leq 1000$ & & 60 & 300 & 1000 \\
 \hline
$M_W$~(GeV) & $79.91\pm 0.39$ & 79.95 & 80.10 & 80.10 & 80.13 \\
$ \Gamma_Z $~(MeV) & $2488.0\pm 7.0 $ & 2488.4 & 2489.0 & 2488.9 & 2488.8 \\
$ \Gamma_{b\bar{b}} $~(MeV) & $383.0\pm 6.0 $ & 379.4 & 377.4 & 376.5 & 375.4
\\
$ \Gamma_{l\bar{l}} $~(MeV) & $83.52\pm 0.28 $ & 83.47 & 83.53 & 83.53
& 83.63 \\
$ \Gamma_{had} $~(MeV) & $1739.9\pm 6.3 $ & 1740.3 & 1738.8 & 1738.2 & 1737.7
\\
$R(\Gamma_{b\bar{b}}/\Gamma_{had})$ & $0.220 \pm 0.003$ & 0.218 & 0.217 &
0.217 & 0.216 \\
$R(\Gamma_{had}/\Gamma_{l\bar{l}})$ & $20.83 \pm 0.06$ & 20.85 & 20.82 &
20.81 & 20.78 \\
$\sigma _h^P (nb)$ & $41.45 \pm 0.17 $ & 41.41 & 41.37 & 41.38 & 41.40 \\
$ g_V $ & $-0.0372 \pm 0.0024 $ & -0.0372 & -0.0341 & -0.0334 & -0.0334 \\
$ g_A $ & $-0.4999 \pm 0.0009 $ & -0.5000 & -0.5003 & -0.5005 & -0.5006 \\
\hline
$\chi^2 $ & & 0.9621 & 4.2664 & 6.0802 & 7.7723 \\
\hline
$ \sin ^2{\theta_W}$ & 0.2321 & 0.2314 & 0.2284 & 0.2283 & 0.2278 \\
\hline
$ \Delta r $ & 0.0623 & 0.06022 & 0.05162 & 0.05131 & 0.04967 \\
\hline \hline
\end{tabular}
\caption{Numerical results including full EWRC for
nine experimental parameters of the Z-decay and $M_W$. The results of QBA
are shown also for comparison.
Each pair of $m_t$ and $m_H$ represents the case of the best $\chi ^2$
-fit to the 1993 LEP data and $M_W = 79.91(39)$ GeV.}
\end{center}
\end{table}
\begin{table}
\begin{center}
\begin{tabular}{|c||c|c||c|c|c|} \hline \hline
 & Experiment & QBA & Full EW & Full EW & Full EW \\
\hline
 $m_t$~(GeV) & 150  & & 126 & 142 & 160 \\
 $ m_H$~(GeV) & 60 $\leq m_H \leq 1000$ & & 60 & 300 & 1000 \\
 \hline
$M_W$~(GeV) & $80.22\pm 0.26$ & 79.95 & 80.13 & 80.13 & 80.15 \\
$ \Gamma_Z $~(MeV) & $2488.0\pm 7.0 $ & 2488.4 & 2490.2 & 2489.7 & 2489.3 \\
$ \Gamma_{b\bar{b}} $~(MeV) & $383.0\pm 6.0 $ & 379.4 & 377.3 & 376.4 & 375.3
\\
$ \Gamma_{l\bar{l}} $~(MeV) & $83.52\pm 0.28 $ & 83.47 & 83.53 & 83.63
& 83.63 \\
$ \Gamma_{had} $~(MeV) & $1739.9\pm 6.3 $ & 1740.3 & 1739.6 & 1738.8 & 1738.1
\\
$R(\Gamma_{b\bar{b}}/\Gamma_{had})$ & $0.220 \pm 0.003$ & 0.218 & 0.217 &
0.216 & 0.216 \\
$R(\Gamma_{had}/\Gamma_{l\bar{l}})$ & $20.83 \pm 0.06$ & 20.85 & 20.83 &
20.79 & 20.78 \\
$\sigma _h^P (nb)$ & $41.45 \pm 0.17 $ & 41.41 & 41.38 & 41.39 & 41.40 \\
$ g_V $ & $-0.0372 \pm 0.0024 $ & -0.0372 & -0.0344 & -0.0337 & -0.0335 \\
$ g_A $ & $-0.4999 \pm 0.0009 $ & -0.5000 & -0.5004 & -0.5006 & -0.5007 \\
\hline
$ \chi^2 $ & & 2.0545 & 4.1346 & 5.9538 & 7.5517 \\
\hline
$\sin^2 {\theta_W}$ & 0.2261 & 0.2314 & 0.2278 & 0.2279 & 0.2275 \\
\hline
$ \Delta r $ & 0.0448 & 0.06022 & 0.04975 & 0.04991 & 0.04895 \\
\hline \hline
\end{tabular}
\caption{The same as Table 1 but for the experimental
$M_W = 80.22 \pm 0.26$ GeV.}
\end{center}
\end{table}
%
%
\begin{table}
\begin{center}
\begin{tabular}{|c||c|c||c|c|c|} \hline \hline
 & Experiment & QBA & Full EW & Full EW & Full EW \\
  \hline
 $m_t$~(GeV) & $174\pm 10^{+13}_{-12}$  & & 187 & 172 & 155 \\
 $ m_H$~(GeV) & 60 $\leq m_H \leq 1000$ & & 1000 & 300 & 60 \\
 \hline
$M_W$~(GeV) & $80.38\pm 0.23$ & 79.95 & 80.33 & 80.32 & 80.30 \\
$ \Gamma_Z $~(MeV) & $2497.1\pm 3.8 $ & 2488.7 & 2496.8 & 2497.3 & 2496.7 \\
$ \Gamma_{l\bar{l}} $~(MeV) & $83.98\pm 0.18 $ & 83.49 & 83.90 & 83.87
& 83.80 \\
$ \Gamma_{had} $~(MeV) & $1746.0\pm 4.0 $ & 1740.5 & 1743.4 & 1744.1 & 1744.2
\\
$R(\Gamma_{b\bar{b}}/\Gamma_{had})$ & $0.2210 \pm 0.0019$ & 0.2180 & 0.2149 &
0.2155 & 0.2160 \\
$R(\Gamma_{had}/\Gamma_{l\bar{l}})$ & $20.790 \pm 0.04$ & 20.847 & 20.778 &
20.794 & 20.813 \\
$\sigma _h^P (nb)$ & $41.51 \pm 0.12 $ & 41.41 & 41.41 & 41.40 & 41.39 \\
$ g_V / g_A $ & $0.0711 \pm 0.002 $ & 0.0745 & 0.0711 & 0.0714 & 0.0723 \\
\hline
$\chi^2 $ & & 25.8 & 11.6 & 9.99 & 9.86 \\
\hline
$ \sin ^2{\theta_W}$ & 0.2231 & 0.2314 & 0.2240 & 0.2242 & 0.2245 \\
\hline
$ \Delta r $ & 0.0355 & 0.06022 & 0.03841 & 0.03913 & 0.03998 \\
\hline \hline
\end{tabular}
\caption{Numerical results including full EWRC for
seven experimental parameters of the Z-decay and $M_W$. The case of QBA
is shown also for comparison.
Each pair of $m_t$ and $m_H$ represents the case of the best $\chi ^2$-
fit to the 1994 LEP data and $M_W = 80.38(23)$ GeV.}
\end{center}
\end{table}
%
%
\begin{table}
\begin{center}
\begin{tabular}{|c||c|c||c|c|c|} \hline \hline
 & Experiment & QBA & Full EW & Full EW & Full EW \\
\hline
 $m_t$~(GeV) &  $174\pm 10^{+13}_{-12}$  & & 185 & 174 & 153 \\
 $ m_H$~(GeV) & 60 $\leq m_H \leq 1000$ & & 1000 & 400 & 60 \\
 \hline
$M_W$~(GeV) & $80.21\pm 0.16$ & 79.95 & 80.32 & 80.31 & 80.29 \\
$ \Gamma_Z $~(MeV) & $2497.1\pm 3.8 $ & 2488.7 & 2496.3 & 2496.8 & 2496.2 \\
$ \Gamma_{l\bar{l}} $~(MeV) & $83.98\pm 0.18 $ & 83.49 & 83.88 & 83.87
& 83.79 \\
$ \Gamma_{had} $~(MeV) & $1746.0\pm 4.0 $ & 1740.5 & 1743.0 & 1743.6 & 1743.9
\\
$R(\Gamma_{b\bar{b}}/\Gamma_{had})$ & $0.2210 \pm 0.0019$ & 0.2180 & 0.2150 &
0.2154 & 0.2161 \\
$R(\Gamma_{had}/\Gamma_{l\bar{l}})$ & $20.790 \pm 0.04$ & 20.847 & 20.779 &
20.791 & 20.813 \\
$\sigma _h^P (nb)$ & $41.51 \pm 0.12 $ & 41.41 & 41.41 & 41.40 & 41.38 \\
$ g_V / g_A $ & $0.0711 \pm 0.002 $ & 0.0745 & 0.0707 & 0.0711 & 0.0720 \\
\hline
$ \chi^2 $ & & 25.8 & 12.1 & 10.6 & 10.1 \\
\hline
$\sin^2 {\theta_W}$ & 0.2263 & 0.2314 & 0.2243 & 0.2244 & 0.2247 \\
\hline
$ \Delta r $ & 0.0455 & 0.06022 & 0.03925 & 0.03957 & 0.04070 \\
\hline \hline
\end{tabular}
\caption{The same as Table 3 but for the experimental
$M_W = 80.21 \pm 0.16$ GeV.}
\end{center}
\end{table}
%

Numerical results for the best $\chi^2$ fit to the 1993 LEP experimental
parameters of $Z$-decay
are shown in Tables 1 and 2
for $M_W = 79.91(39) $ GeV and $ M_W = 80.22(26)$ GeV respectively
as experimental inputs.
The results for the best $\chi^2$-fit to the improved 1994 LEP data
and $M_W$ are
given in Tables 3 and 4 for $M_W = 80.38(23)$ GeV and $M_W = 80.21(16)$ GeV
respectively.
Also included in the Tables are the results of QBA
as well as the output $\sin^2 \theta_W$ and $\Delta r$
for comparison.
We see that the contributions of the weak corrections are
generally small and in particular for the 1993 data the QBA
is close to the experimental values
within the uncertainty of the measurements.

The near absence of the pure weak loop contributions to the
radiative corrections for the 1993 data is more impressive for $M_W = 79.91$
GeV than for $M_W = 80.22$ GeV. At closer examination, however, the QBA
in the latter case over-estimates the radiative corrections and the full
one-loop EWRC fair better.

{}From the global fit to the data with two variables $m_t$ and $m_H$
in the range $60-1000 $ GeV, we find the best fits to
the 1993 data is obtained by $m_t = 142^{-16}_{+18} $ GeV, the central value
being the best $\chi^2$ case of $m_H = 300$ GeV in Table 2.
The best global fits to the 1993 data give a rather
stable output $M_W = 80.13\pm 0.03~$ GeV if the full EWRC are
taken into account, which
is to be contrasted to the output $M_W = 79.95$ GeV from the QBA,
for either experimental $M_W$ value. Also
$\sin^2 \theta_W = 0.2279\pm 0.0005$ in the case of the full EWRC is
to be compared to $\sin ^2\theta _W = 0.2314$
in the case of QBA.
While the 1993 world average value of
$M_W$ supports strongly for the evidence of the full EWRC
in the LEP data, the QBA appears to be in statistically comparable
agreement, i.e., within $2\sigma$, with the precisions of the 1993 data.
If $M_W$ were to be definitely at around 79.95 GeV with the uncertainty
of the 1993 LEP data,
then the QED correction would have been all that was observed at LEP and
one would have been cultivating the null result of the weak correction
to produce the
range of t-quark mass as pointed out in [7].

The situation with the $\chi^2$-fit to the improved 1994 LEP data and
$M_W$ is significantly
different from the case of the 1993 data as one can see from Tables 3 and 4.
Not only there is clear evidence for the full EWRC in each of the seven
LEP data but also the QBA gives distinctively inferior $\chi^2$ in
either case of new $M_W$.
{}From the best fits to the 1994 data, one gets again a stable output
$M_W = 80.31 \pm 0.02 $ GeV for $m_H$ in the range of 60 - 1000 GeV.
In particular, the CDF $m_t$ value $174$ GeV is a possible output solution
(in the case $M_W = 80.21(16) $ GeV ) but with a $m_H $ about 400  GeV
among the many possible combinations of $(m_t, m_H)$ given by the 'Best.fit'
 curve in Fig. 1.
In general the $\chi^2$-value tends to prefer lower $m_t$ and accordingly
 smaller $m_H$
combination of the curve but any pair of $(m_t, m_H)$ on this curve is
 statistically
comparable to each other.
We see from Fig. 1 that the best global fits to the 1994
 data are obtained
by $m_t = 153 - 185$ GeV for $m_H = 60 - 1000$ GeV.

Fig.2 shows how $M_W$ changes with $m_t$ for fixed $m_H$ from
full EWRC where the new world average $M_W$ and CDF $m_t$ are also shown.
The central values of the world average  $M_W$ and CDF $m_t$ are
 consistent with a Higgs scalar mass somewhat heavier than 1000 GeV,
though $m_H = 200$ GeV is only less than 1.5 $\sigma$ away.
Clearly a better precision measurement of $M_W$ is desired to distinguish
different $m_H$. For example, a change of $m_H$ by 200 GeV, i.e.,
from 400 GeV to 200 GeV, requires from the best $\chi^2$-fits a change
of 9 GeV in $m_t$, i.e., from 174 GeV to 165 GeV, as one can see from Fig. 1.
This in turn requires a precision of 20 MeV or better in $M_W$ from Fig. 2.

In short, we find that the QBA
is in agreement with the 1993 data within $2\sigma$ level of
accuracy but the new world average value of $M_W$ and the improved
1994 LEP data disfavor the QBA and definitely support for the
evidence of the nonvanishing weak-loop correction.
Furthermore, the CDF $m_t$ is a solution of the minimal $\chi^2$-fit to
the 1994 data but
then the Higgs scalar mass is {\it about} 400 GeV.
Further precision measurement of $M_W$ can provide
a real test of the standard model as it will give a tight constraint for the
needed amount of the EWRC and provide a profound implication to the mass of
t-quark and Higgs scalar in the context of the standard model.
 If $M_W$
is determined within 20 MeV uncertainty, $\Delta r$ within the
context of the standard model
can distinguish the mass range of the t-quark and Higgs scalar and
provide a crucial test for and even the need of new physics beyond the
standard model.


\section*{References}
\item[1.] See, for example, L. Rolandi, in: Proc. XXVI ICHEP 1992, CERN-PPE/92
          -175 (1992); M. P. Altarelli, talk given at Les Rencontres
          de Physique de la Vallee d'Aoste (La Thuil, 1993),
          LNF-93/019(p); and J. Lefranceis, in: Proc.
          Int. EPS. Conf. H. E. Phys. (Marseille, July 1993).
\item[2.] P. Clarke; Y. K. Kim; B. Pictrzyk; P. Siegris; and M. Woods, in:
          Proc. 29th Rencontres de Moriond (Meribel, 1994), and D. Schaile (
          private communication).
\item[3.] F. Dydak, in: Proc. 25 Int. Conf. H. E. Phys.,
          Eds. K. K. Phua, Y. Yamaguchi (World Scientific, Singapore, 1991).
\item[4.] W. J. Marciano, Phys. Rev. {\bf D20} (1979) 274;
          A. Sirlin, Phys. Rev. {\bf D22} (1980) 971; (1984) 89;
          A. Sirlin and W. J. Marciano, Nucl. Phys. {\bf B189} (1981) 442;
          and A. Sirlin, NYU-TH-93/11/01. See also W. Hollik, in:
          {\it Precision Tests of the Standard Model}, ed. P.
          Langacker (World Scientific Pub., 1993) and G. Altarelli, in: Proc.
          Int.EPS. Conf. H. E. Phys. (Marseille, July 1993).
\item[5.] D. Saltzberg, Fermilab-Conf-93/355-E, and  S. B. Kim (private
          communication)
\item[6.] The most recent $m_t$ lower bound from D0 is 131 GeV,
           S. Abachi et al, Phys. Rev. Lett. {\bf 72} (1994) 2138 while
           the CDF reported $m_t = 174 \pm 10^{+13}_{-12}$ GeV in F.
           Abe et al., FERMILAB-PUB-94/097-E, and Phys. Rev. Lett. {\bf
           73} (1994) 255.
\item[7.] V. A. Novikov, L. B. Okun and M. I. Vysotsky, Mod. Phys. Lett.
          {\bf A8} (1993) 5929. See, however, CERN-TH-7214/94 for more recent
          analysis.
\item[8.] D. Bardin et al., CERN-TH-6443-92 (1992).
\item[9.] See, for example, W. Hollik (Ref. 9); CERN Yellow Book CERN 89-08,
	   vol.1, p45 ; and K. Kang, in: Proc.
           14th Int. Workshop Weak Interactions and Neutrinos (Seoul,
           1993), Brown-HET-931 (1993).
\item[10.] Particle Data Group, Review of Particle Properties,
          Phys. Rev. {\bf D45}, No.11, Part II (1992).
\item[11.] CDF Collab., F. Abe et al., Phys. Rev. {\bf D43} (1991) 2070.
\item[12.] Kyungsik Kang and Sin Kyu Kang, SNUTP-94-59 (to be published).
\item[13.] T. Hebbeker, Aachen preprint PITHA 91-08 (1991); and
           S. G. Gorishny, A. L. Kataev and S. A. Larin, Phys. Lett. {\bf B259}
           (1991) 144. See also L. R. Surguladze and M. A. Samuel,
           Phys. Rev. Lett. {\bf 66} (1991) 560.
\item[14.] M. Consoli and W. Hollik, in {\it Z Physics at LEP 1}, Vol. 1,
           eds. G. Altarelli et al., CERN 89-08 (1989).

\section*{Figure Captions}
\item[Fig. 1]: The mass ranges of $m_t$ and $m_H$ from the minimal $\chi^2$-fit
 to the 1994 LEP data and $M_W$ = 80.21 GeV.
\item[Fig. 2]: $M_W$ versus $m_t$ for fixed values of $m_H$ from the full
 radiative correction in the standard model.
 The case of the minimal $\chi^2$-fit to the 1994 LEP data corresponding
 to the full EWRC in Table 4 are indicated by $\Diamond $.
\end{description}
\end{document}